\documentstyle[12pt]{article}
\catcode`\@=11
\def\numberbysection{\@addtoreset{equation}{section}
\def\theequation{\thesection.\arabic{equation}}}
\numberbysection

\newcommand{\abs}[1]{\vert#1\vert}
\newcommand{\as}{{\mathrm as}}
\newcommand{\bas}{{_{\vphantom{M}}}}
\newcommand{\beq}{\begin{equation}}
\newcommand{\beqa}{\begin{eqnarray}}
\newcommand{\cotanh}{\mathop{\mathrm cotanh}}
\renewcommand{\d}{{\mathrm d}}
\newcommand{\de}{{\mathrm def}}
\newcommand{\dpar}{\partial}
\newcommand{\e}{{\mathrm e}}
\newcommand{\eeq}{\end{equation}}
\newcommand{\eeqa}{\end{eqnarray}}
\newcommand{\eq}{{\mathrm eq}}
\newcommand{\erf}{\mathop{\mathrm erf}}
\newcommand{\erfc}{\mathop{\mathrm erfc}}
\newcommand{\fd}{fluc\-tu\-a\-tion-dis\-si\-pa\-tion }
\newcommand{\frad}[2]{{\displaystyle{\displaystyle#1\over\displaystyle#2}}}
\newcommand{\g}{\gamma}
\newcommand{\haut}{{^{\vphantom{M}}}}
\renewcommand{\i}{{\mathrm i}}

\newcommand{\mean}[1]{\left\langle#1\right\rangle}
\newcommand{\ps}{p_s}
\newcommand{\s}{\sigma}
\newcommand{\un}[1]{\noindent $\bullet$ {\it#1}}
\newcommand{\A}{{\mathrm A}}

\newcommand{\F}{{\mathrm F}}
\renewcommand{\H}{{\cal H}}
\renewcommand{\L}{{\mathrm L}}
\newcommand{\Ls}{{\mathrm L}_s}
\newcommand{\R}{\rho}
\newcommand{\X}{{\cal X}}
\begin{document}
\centerline{\Large\bf Response of non-equilibrium systems at criticality:}
\vspace{.3cm}
\centerline{\Large\bf Exact results for the Glauber-Ising chain}
\vspace{1cm}
\centerline{\large
by C.~Godr\`eche$^{a,}$\footnote{godreche@spec.saclay.cea.fr}
and J.M.~Luck$^{b,}$\footnote{luck@spht.saclay.cea.fr}}
\vspace{1cm}
\centerline{$^a$Service de Physique de l'\'Etat Condens\'e,
CEA Saclay, 91191 Gif-sur-Yvette cedex, France}
\vspace{.1cm}
\centerline{$^b$Service de Physique Th\'eorique,
CEA Saclay, 91191 Gif-sur-Yvette cedex, France}
\vspace{1cm}
\begin{abstract}
We investigate the non-equilibrium two-time correlation
and response functions and the associated \fd ratio
for the ferromagnetic Ising chain with Glauber dynamics.
The scaling behavior of these quantities at low temperature
and large times is studied in detail.
This analysis encompasses the self-similar domain-growth (aging) regime,
the spatial and temporal Porod regimes, and the convergence toward equilibrium.
The \fd ratio admits a non-trivial limit value
$X_\infty=1/2$ at zero temperature, and more generally in the aging regime.
\end{abstract}
\vfill
\noindent To appear in Journal of Physics A
\hfill S/99/047
\vskip -6pt
\noindent P.A.C.S.: 05.70.Ln, 64.60.Ht, 05.40.+j
\hfill T/99/107
\newpage
\section{Introduction}

The ferromagnetic Ising chain with Glauber dynamics~\cite{glauber}
is one of the simplest non-equilibrium systems.
Consider a finite system consisting of $N$ Ising spins $\s_n=\pm1$,
with Hamiltonian
\beq
\H=-J\sum_n\s_n\s_{n+1}.
\label{ham}
\eeq
In its heat-bath formulation Glauber dynamics consists in
picking, at every time step $\delta t=1/N$, a site $n=1,\dots,N$ at random,
and updating its spin $\s_n(t)$ according to the stochastic rule
\beq
\s_n(t)\to\left\{\matrix{
+1&\mbox{with prob.}\,\frad{1+\tanh(\beta h_n(t))}{2},\hfill\cr\cr
-1&\mbox{with prob.}\,\frad{1-\tanh(\beta h_n(t))}{2\bas},\hfill\cr}\right.%}
\label{update}
\eeq
where $\beta=1/T$ is the inverse temperature,
and the local field $h_n(t)$ acting on $\s_n(t)$ reads
\beq
h_n(t)=J(\s_{n-1}(t)+\s_{n+1}(t)).
\eeq
In the thermodynamic limit $(N\to\infty)$,
each site is thus updated according to a Poisson process.

At positive temperature, starting from a random initial condition,
obtained for instance by a quench from infinitely high temperature,
the system relaxes to its paramagnetic equilibrium state.
The situation at zero temperature is different.
Indeed, still starting from a random initial condition,
the system is unable to relax to any of its two
ferromagnetically ordered, symmetry-related, equilibrium states.
On the contrary, domains of positive and negative magnetization grow forever,
and, in the scaling regime, the system becomes statistically self-similar
with only one characteristic length scale, the mean size of domains.
This coarsening process is a consequence of the existence
of spontaneous symmetry breaking~\cite{langer,bray1}.

The Ising chain is special
because its critical temperature $T_c$ is equal to zero.
However its dynamical behavior illustrates that of generic
statistical-mechanical models in the absence of quenched disorder.
For the latter, starting from a random initial condition,
the system relaxes exponentially to equilibrium
in the high-temperature phase $(T>T_c)$.
At equilibrium, two-time quantities such as the
correlation function $C(t,s)$ or the response function $R(t,s)$,
where $s$~(waiting time) is smaller that $t$~(observation time),
only depend on the time difference
\beq
\tau=t-s\ge0,
\label{tau}
\eeq
and they are simply related to each other by the \fd theorem:
\beq
R_\eq(\tau)=-\beta\,\frac{\d C_\eq(\tau)}{\d\tau}.
\label{fdt}
\eeq
In the low-temperature phase $(T<T_c)$ the system undergoes phase ordering.
In this non-equilibrium situation, $C(t,s)$ and $R(t,s)$
are non-trivial functions of both time variables,
which only depend on their ratio at late times,
i.e., in the self-similar scaling regime.
This behavior is usually referred to as aging~\cite{revue}.
Moreover, no such simple relation as eq.~(\ref{fdt}) holds
between correlation and response, i.e.,
$R(t,s)$ and $\dpar C(t,s)/\dpar s$ are no longer proportional.
It is then natural to characterize the distance to equilibrium
of an aging system by the so-called \fd ratio~\cite{revue,ckp,ck}
\beq
X(t,s)=\frac{T\,R(t,s)}{\,\frad{\dpar C(t,s)\haut}{\dpar s}\,}.
\label{dx}
\eeq

In recent years, several works~\cite{revue,ckp,ck,x1,x2,autres}
have been devoted to the study of the \fd ratio.
In the low-temperature phase of aging systems,
such as glasses and spin glasses, or of systems exhibiting domain growth,
$X(t,s)$ turns out to be a non-trivial function of its two arguments.
In the case of coarsening systems,
analytical and numerical studies indicate that the limit \fd ratio,
\beq
X_\infty=\lim_{s\to\infty}\lim_{t\to\infty}X(t,s),
\label{xinfdef}
\eeq
vanishes throughout the low-temperature phase~\cite{x1,x2}.

However, to date, only very little attention has been devoted
to the \fd ratio $X(t,s)$ for non-equilibrium systems {\it at criticality}.
From now on, we will only have in mind non-disordered systems.
For instance one may wonder whether there exists,
for a given model, a well-defined limit $X_\infty$ at $T=T_c$,
different from its value in the low-temperature phase.
Indeed, a priori, for a system without disorder, such as a ferromagnet,
quenched from infinitely high temperature to its critical point,
the limit \fd ratio $X_\infty$ at $T=T_c$
(if it exists) may take any value between $X_\infty=1$ ($T>T_c$: equilibrium)
and $X_\infty=0$ ($T<T_c$: domain growth).

The only cases of critical systems for which the \fd ratio has been
considered are, to our knowledge, the models of ref.~\cite{ckp}
(random walk, free Gaussian field,
and two-dimensional X-Y model at zero temperature)
which share the limit \fd ratio $X_\infty=1/2$,
and the backgammon model, a mean-field model for which $T_c=0$,
where it has been shown that $X_\infty=1$,
up to a large logarithmic correction,
for both energy fluctuations and density fluctuations~\cite{fr97,gl99}.

In the present work we investigate the non-equilibrium
response function and \fd ratio for the Glauber-Ising chain,
both at zero temperature and in the scaling regime of low temperatures.
Exact results for these two-time quantities are derived,
exploiting the solvability of the model.
We then perform a detailed analysis of their scaling
behavior at low temperature and large times.
Since the computation of $R(t,s)$ and of $X(t,s)$
requires the knowledge of the equal-time correlation function,
of the two-time correlation $C(t,s)$, and of its derivative with respect to one
of the two times~[see eq.~(\ref{dx})], we will include
the computation and the scaling analysis of the latter quantities,
though a number of papers have already
dealt with the study of correlations in the Glauber-Ising
chain~\cite{glauber,cox,bray2,amar,bray3,PBS}.
Our presentation will thus be complete and self-contained,
and the computation of all quantities will follow
the same integral-transform methods.

In a forthcoming publication~\cite{ustocome},
we will present a study of the critical response and \fd ratio
for the ferromagnetic spherical model in any dimension $d>2$,
and for the two-dimensional Ising model.
Both models possess a whole low-temperature phase for $T<T_c$,
hence they are faithful representatives of generic domain-growth systems.

One salient outcome of these joint works is the realization
that the limit \fd ratio $X_\infty$
is a novel universal characteristic of critical dynamics
(see the discussion at the end of the present paper).
For the Ising chain at its critical temperature $T_c=0$,
we obtain the value $X_\infty=1/2$.
The occurrence of a common value for $X_\infty$ between the Ising chain
and the models considered in ref.~\cite{ckp} seems rather coincidental,
for the time being.

In the following, we shall make use of the parameters $\mu$ and
$\g$, defined by
\beq
\mu=-\ln\tanh(\beta J),\qquad\g=\tanh(2\beta J)=\frac{1}{\cosh\mu}.
\eeq
These parameters are related to the equilibrium correlation length
$\xi_\eq$~\cite{baxter} and to the equilibrium relaxation time $\tau_\eq$
[see eq.~(\ref{relax})] by
\beq
\xi_\eq=\frac{1}{\mu},\qquad\tau_\eq=\frac{1}{1-\g}.
\label{xitau}
\eeq
Both diverge exponentially fast as the critical temperature $T_c=0$ is
approached, according to
\beq
\xi_\eq\approx\frac{\e^{2\beta J}}{2},
\qquad\tau_\eq\approx\frac{2}{\mu^2}\approx\frac{\e^{4\beta J}}{2},
\eeq
hence we have the scaling law
\beq
\tau_\eq\approx 2\xi_\eq^2,
\label{tauxi}
\eeq
corresponding to a dynamical critical exponent $z=2$,
and reflecting the diffusive nature of domain growth at zero temperature.

We close up this introduction
with a discussion about relevant time scales.
For two-time quantities such as $C(t,s)$, $R(t,s)$, $X(t,s)$,
three time scales are to be compared in the scaling regime
of long times and low temperature, namely $s$, $\tau=t-s$, and $\tau_\eq$.
Considering any two of these time scales as comparable,
with an arbitrary ratio between them,
and small (or large) compared to the third one,
six different regimes can be defined a priori.
Three of them will be of interest in this work,
which we summarize here for convenience:
\beq
\tau\sim\tau_\eq\ll s:\quad\mbox{Equilibrium}
{\hskip 45.6mm}
\label{l1}
\eeq
\beq
s\sim\tau\ll\tau_\eq:\quad\mbox{Self-similar domain growth (aging)}
{\hskip 4.45mm}
\label{l2}
\eeq
\beq
\tau\ll s\sim\tau_\eq:\quad\mbox{Early-time or temporal Porod regime}
\label{l3}
\eeq
Aging persists forever at zero temperature,
just as the coarsening process itself,
while it is interrupted at the time scale $\tau_\eq$
at any low but finite temperature.

\section{Magnetization profile}

In this section we present the methods used throughout the paper
on the example of the relaxation of the magnetization profile
\beq
M_n(t)=\mean{\s_n(t)},
\eeq
where the brackets denote an average over the thermal history of the system
and over the initial conditions, which are unspecified for the time being.

The time evolution of the magnetization $M_n(t)$
is readily deduced from the stochastic rule~(\ref{update}), and is given by
\beq
\frac{\d M_n(t)}{\d t}=-M_n(t)+\mean{\tanh\haut(\beta h_n(t))}.
\label{mvt}
\eeq
Since the local field $h_n$ only assumes three symmetric values, $0$ and
$\pm2J$, we have
\beq
\tanh(\beta h_n)=\frac{\g}{2J}\,h_n=\frac{\g}{2}(\s_{n-1}+\s_{n+1}),
\eeq
leading to the linear evolution equations
\beq
\frac{\d M_n(t)}{\d t}=-M_n(t)+\frac{\g}{2}(M_{n-1}(t)+M_{n+1}(t)).
\label{mdot}
\eeq
The existence of closed, linear equations for the magnetization,
and more generally for higher-order correlation functions, ensures the
solvability of the model~\cite{glauber}.

In order to solve coupled equations of the form~(\ref{mdot}),
we shall make an extensive use of Laplace and Fourier transforms.
For any quantity $f_n(t)$, depending both on continuous time $t$
and on the discrete site label $n$, we introduce

\un{\null} the temporal Laplace transform
\beq
f^\L_n(p)=\int_0^\infty f_n(t)\,\e^{-pt}\,\d t,\qquad
f_n(t)=\int\frac{\d p}{2\pi\i}\,f^\L_n(p)\,\e^{pt},
\label{lapdef}
\eeq

\un{\null} the spatial Fourier transform
\beq
f^\F(q,t)=\sum_n f_n(t)\,\e^{-\i nq},\qquad
f_n(t)=\int_0^{2\pi}\frac{\d q}{2\pi}\,f^\F(q,t)\,\e^{\i nq},
\eeq

\un{\null} the Fourier-Laplace transform
\beq
f^{\F\L}(q,p)=\sum_n\int_0^\infty f_n(t)\,\e^{-(pt+\i nq)}\,\d t,
\qquad f_n(t)=\int\frac{\d p}{2\pi\i}
\int_0^{2\pi}\frac{\d q}{2\pi}\,f^{\F\L}(q,p)\,\e^{pt+\i nq}.
\eeq

Using the above integral transforms, eq.~(\ref{mdot}) can be recast as
\beq
\frac{\d M^\F(q,t)}{\d t}=(\g\cos q-1)M^\F(q,t),
\eeq
or else as
\beq
p\,M^{\F\L}(q,p)=(\g\cos q-1)M^{\F\L}(q,p)+M^\F(q,t=0),
\label{mdotfl}
\eeq
yielding
\beq
M^{\F\L}(q,p)=\frac{M^\F(q,t=0)}{p+1-\g\cos q}.
\label{mfl}
\eeq

Consider first the locally magnetized initial condition
where the spin at the origin is pointing upwards, i.e., $\s_0(t=0)=+1$,
but the configuration is otherwise totally random.
The corresponding solution $G_n(t)$ to eq.~(\ref{mdot})
is the Green's function of the problem.
We have $G_n(t=0)=\delta_{n,0}$, so that $G^\F(q,t=0)=1$.
Eq.~(\ref{mfl}) then reads
\beq
G^{\F\L}(q,p)=\frac{1}{p+1-\g\cos q},
\label{mfl1}
\eeq
hence
\beq
G^\F(q,t)=\e^{(\g\cos q-1)t},
\eeq
and
\beq
G_n(t)=\e^{-t}\int_0^{2\pi}\frac{\d q}{2\pi}\,\e^{\g t\cos q+\i nq}
=\e^{-t}I_n(\g t),
\label{gnt}
\eeq
where $I_n$ is the modified Bessel function.
Eq.~(\ref{mfl1}) also yields
\beq
G_n^\L(p)=\frac{1}{\sqrt{(p+1)^2-\gamma^2}}
\left(\frac{p+1-\sqrt{(p+1)^2-\gamma^2}}{\gamma}\right)^{\abs{n}}.
\label{gnp}
\eeq

Coming back to eq.~(\ref{mdot}), its general solution in direct space
is obtained by inverting eq.~(\ref{mfl}),
yielding the following spatial convolution
\beq
M_n(t)=M_n(t=0)*G_n(t)=\sum_m M_m(t=0)G_{n-m}(t).
\eeq
In particular, for a uniformly magnetized system,
i.e., $M_n(t=0)=M$ for all $n$, we have an exact exponential relaxation
\beq
\frac{M_n(t)}{M}=\sum_n G_n(t)=G^\F(q=0,t)=\e^{-t/\tau_\eq}
\label{relax}
\eeq
at any finite temperature, where the relaxation time $\tau_\eq$
is given by eq.~(\ref{xitau}).

Throughout the following, we shall be mostly interested in the
non-equilibrium scaling regime
of long times, large distances, and low temperatures,
such that $t$ and $\tau_\eq$ are simultaneously large but comparable,
i.e., their ratio $t/\tau_\eq$ is arbitrary,
and the same holds for $n$ and $\xi_\eq$:
\beq
t\sim\tau_\eq\gg1,\qquad n\sim\xi_\eq\gg1.
\label{scadef}
\eeq
The scaling law~(\ref{tauxi}) then implies
\beq
t\sim\tau_\eq\sim n^2\sim\xi_\eq^2.
\label{scafull}
\eeq
As a consequence, $p\sim q^2\sim\mu^2$ are simultaneously small,
and eq.~(\ref{mfl1}) simplifies to
\beq
G^{\F\L}(q,p)\approx\frac{2}{2p+q^2+\mu^2}.
\eeq
Performing the inverse Laplace transform first, we obtain
\beq
G_n(t)\approx\frac{1}{\sqrt{2\pi t}}
\,\exp\left(-\frac{t}{\tau_\eq}-\frac{n^2}{2t}\right),
\label{gnscal}
\eeq
which is the scaling form of the Green's function~(\ref{gnt}),
involving the variables $t/\tau_\eq$ and $n/\sqrt{t}$,
in agreement with eq.~(\ref{scafull}).
At zero temperature
the exponential damping factor $\exp(-t/\tau_\eq)$ is absent,
so that $G_n(t)$ is a function of $n/\sqrt{t}$ only,
reflecting the statistical self-similarity of the coarsening process.
In this case the Gaussian profile is normalized,
emphasizing again the underlying diffusive mechanism.

\section{Equal-time correlation function}

The equal-time correlation function between any two spins is defined as
\beq
C(m,n,t)=\mean{\s_m(t)\s_n(t)}.
\eeq

In the following, unless otherwise specified,
we shall consider a random initial condition,
defined by averaging over all possible initial configurations
with equal weights.
This procedure corresponds to quenching the system at time $t=0$
from its equilibrium state at infinitely high temperature.
Because the invariance under spatial translations along the chain
is preserved by the dynamics,
the correlation function $C(m,n,t)$ only depends on the distance
$\abs{n-m}$ between both spins.
We denote it as
\beq
C(m,n,t)=C_{n-m}(t).
\eeq
We have in particular
\beq
C_0(t)=1.
\label{cond0}
\eeq
The equal-time correlation function can be shown,
in analogy with eq.~(\ref{mdot}),
to obey coupled linear differential equations of the form~\cite{glauber}
\beq
\frac{\d C_n(t)}{\d t}=-2C_n(t)+\g(C_{n-1}(t)+C_{n+1}(t))\qquad(n\ne0),
\label{cdot}
\eeq
with the condition~(\ref{cond0}),
and the initial value $C_n(t=0)=\delta_{n,0}$.

In order to solve eq.~(\ref{cdot}),
we complete them by the corresponding equation for $n=0$,
with a time-dependent source $v(t)$ in the right-hand side,
to be determined in such a way that the condition~(\ref{cond0}) be fulfilled.
In other words we consider the equations
\beq
\frac{\d C_n(t)}{\d t}=-2C_n(t)+\g(C_{n-1}(t)+C_{n+1}(t))+v(t)\delta_{n,0},
\label{cdot0}
\eeq
which read, in Fourier-Laplace space,
\beq
p\,C^{\F\L}(q,p)=2(\g\cos q-1)C^{\F\L}(q,p)+v^\L(p)+1,
\label{cdotfl}
\eeq
yielding
\beq
C^{\F\L}(q,p)=\frac{v^\L(p)+1}{p+2-2\g\cos q}.
\eeq
The condition~(\ref{cond0}) then reads
\beq
\int_0^{2\pi}\frac{\d q}{2\pi}\,C^{\F\L}(q,p)
=\frac{v^\L(p)+1}{\sqrt{(p+2)^2-4\g^2}}=\frac{1}{p},
\eeq
hence
\beq
v^\L(p)=\frac{\sqrt{(p+2)^2-4\g^2}}{p}-1.
\eeq
We thus obtain the result
\beq
C^{\F\L}(q,p)=\frac{\sqrt{(p+2)^2-4\g^2}}{p(p+2-2\g\cos q)},
\label{cfl}
\eeq
or else
\beq
C_n^\L(p)
=\frac{1}{p}\left(\frac{p+2-\sqrt{(p+2)^2-4\g^2}}{2\g}\right)^{\abs{n}},
\label{cnl}
\eeq
which we now discuss.

\un{Equilibrium}

At any finite temperature, and for long enough times $(t\gg\tau_\eq)$,
the correlation function $C_n(t)$ converges to its equilibrium value,
$C_{n,\eq}=\lim_{p\to0}(p\,C_n^\L(p))$.
From eq.~(\ref{cnl}) we recover the well-known expression~\cite{baxter}
\beq
C_{n,\eq}=\left(\frac{1-\sqrt{1-\g^2}}{\g}\right)^{\abs{n}}=\e^{-\mu\abs{n}}.
\label{cneq}
\eeq

\un{Non-equilibrium}

Let us analyze the non-equilibrium properties in the regime~(\ref{scafull})
of low temperatures and long times.
In this regime the expressions~(\ref{cfl}) and~(\ref{cnl}) scale as
\beq
C^{\F\L}(q,p)\approx\frac{2\sqrt{p+\mu^2}}{p(p+q^2+\mu^2)}
\label{csfl}
\eeq
and
\beq
C_n^\L(p)\approx\frac{\e^{-\abs{n}\sqrt{p+\mu^2}}}{p}.
\label{csnl}
\eeq
Eq.~(A.1) leads to the following explicit scaling form
\beq
C_n(t)\approx\frac{1}{2}\left\{
\e^{\mu\abs{n}}\,\erfc\!\left(\frac{\abs{n}}{2\sqrt{t}}+\mu\sqrt{t}\right)
+\e^{-\mu\abs{n}}\,\erfc\!\left(\frac{\abs{n}}{2\sqrt{t}}-\mu\sqrt{t}\right)
\right\}
\label{cnsca}
\eeq
for the equal-time correlation function,
involving two scaling variables, $\mu^2t\approx2t/\tau_\eq$ and $n/\sqrt{t}$.
The error function $\erf z$
and the complementary error function $\erfc z$ are defined as
\beq
\erf z=\frac{2}{\sqrt{\pi}}\int_0^z\e^{-x^2}\,\d x,\qquad
\erfc z=1-\erf z=\frac{2}{\sqrt{\pi}}\int_z^\infty\e^{-x^2}\,\d x.
\eeq
An alternative expression for $C_n(t)$ is
\beq
C_n(t)\approx\frac{\abs{n}}{2}\int_0^t\frac{\d u}{\sqrt{\pi u^3}}
\exp\left(-\mu^2u-\frac{n^2}{4u}\right),
\eeq
which can be obtained either from eq.~(\ref{csfl}) or from eq.~(\ref{cnsca}).
The following limiting situations are of interest.

\noindent(i)
At zero temperature, and more generally in the
domain-growth regime $(1\ll t\ll\tau_\eq)$,
we can set $\mu=0$ in eq.~(\ref{cnsca}), which thus simplifies to
\beq
C_n(t)\approx\,\erfc\!\left(\frac{\abs{n}}{2\sqrt{t}}\right).
\label{cna}
\eeq
This expression~\cite{bray1,bray3} involves only one scaling variable,
$n/\sqrt{t}$, reflecting the fact that, in this regime,
the pattern formed by the domains is self-similar.

\noindent(ii)
In the opposite regime $(t\gg\tau_\eq)$,
the expression~(\ref{cnsca}) converges exponentially fast
to its equilibrium value~(\ref{cneq}), as
\beq
C_n(t)-C_{n,\eq}
\approx-\frac{\mu\abs{n}}{4\sqrt{2\pi}}\left(\frac{t}{\tau_\eq}\right)^{-3/2}
\,\e^{-2t/\tau_\eq}.
\label{cvcn}
\eeq

\noindent(iii)
The short-distance or Porod regime
\beq
1\ll\abs{n}\ll\xi_\eq\sim\sqrt{t}
\label{porods}
\eeq
is yet another situation of interest~\cite{bray1}.
In this regime, eq.~(\ref{cnsca}) becomes
\beq
C_n(t)\approx 1-A(t)\abs{n},
\label{cporods}
\eeq
with
\beq
A(t)=\frac{\e^{-\mu^2t}}{\sqrt{\pi t}}+\mu\,\erf(\mu\sqrt{t}).
\label{as}
\eeq
This result may also be found from eq.~(\ref{csnl}), by noticing that
$|n|\sqrt{p+\mu^2}\ll1$, and using eq.~(A.2).
The expression~(\ref{as}) for the amplitude $A(t)$
interpolates between the power-law behavior
\beq
A(t)\approx\frac{1}{\sqrt{\pi t}}
\label{adom}
\eeq
in the domain-growth regime, in agreement with eq.~(\ref{cna}),
and the equilibrium value
\beq
A_\eq\approx\mu,
\label{aeq}
\eeq
in agreement with eq.~(\ref{cneq}).

The result~(\ref{cporods}) also holds for finite values
$\abs{n}=0,1,2,\dots$ of the distance between spins,
in the scaling regime $(\xi_\eq\gg1$, $t\gg1)$.
In particular the density of defects (domain walls) in the system reads
\beq
\rho_\de(t)=\frac{1-C_1(t)}{2}\approx\frac{A(t)}{2}.
\eeq
The result~(\ref{cporods}) is thus in quantitative agreement
with the general predictions given in ref.~\cite{bray1}
on the so-called Porod singularities of correlation functions,
whose form depends on the nature of topological defects,
and whose amplitude is known in terms of the density $\rho_\de$
of these defects.

The spatial range over which ferromagnetic order has propagated
at time $t$ can be measured by the dimensionless susceptibility
\beq
\chi(t)=\sum_nC_n(t)=C^\F(q=0,t),
\label{cbc}
\eeq
for which eq.~(\ref{cfl}) yields
\beq
\chi^\L(p)=\frac{1}{p}\sqrt\frac{p+2+2\g}{p+2-2\g}.
\eeq
Throughout the non-equilibrium scaling regime~(\ref{scafull}), we have
\beq
\chi^\L(p)\approx\frac{2}{p\sqrt{p+\mu^2}},
\label{bcfl}
\eeq
i.e., using eq.~(A.3),
\beq
\chi(t)\approx\frac{2}{\mu}\,\erf\left(\mu\sqrt{t}\right).
\label{bcsca}
\eeq
This result interpolates between the square-root behavior
\beq
\chi(t)\approx4\sqrt\frac{t}{\pi}
\label{bc0}
\eeq
in the domain-growth regime $(t\ll\tau_\eq)$,
in agreement with eq.~(\ref{cna}), and the limit
\beq
\chi_\eq\approx\frac{2}{\mu}
\label{ceq}
\eeq
as equilibrium is approached $(t\gg\tau_\eq)$.
This last expression is the scaling behavior of the exact equilibrium value
\beq
\chi_\eq=\cotanh\frac{\mu}{2}
=\frac{\g+1-\sqrt{1-\g^2}}{\g-1+\sqrt{1-\g^2}}
=\e^{2\beta J},
\label{bceq}
\eeq
corresponding to eq.~(\ref{cneq}).

Finally, the dimensionless product
\beq
Q(t)=A(t)\chi(t)=2\,\erf(\mu\sqrt{t})
\left(\frac{\e^{-\mu^2t}}{\mu\sqrt{\pi t}}+\erf(\mu\sqrt{t})\right)
\eeq
is a form factor characteristic of the correlation profile.
It increases from the value $Q=4/\pi$
in the domain-growth (aging) regime~[see eqs.~(\ref{adom}), (\ref{bc0})]
to the equilibrium value $Q_\eq=2$~[see eqs.~(\ref{aeq}), (\ref{ceq})].

\section{Two-time correlation function}

We now consider the two-time correlation function
\beq
C(m,n,t,s)=\mean{\s_n(t)\s_m(s)},
\eeq
where $s$ is the waiting time and $t=s+\tau$ is the observation time.
For a random initial condition, invariance under spatial translations yields
\beq
C(m,n,t,s)=C_{n-m}(t,s).
\eeq

This two-time correlation function can be shown,
again in analogy with eq.~(\ref{mdot}),
to obey the coupled linear partial differential equations~\cite{glauber}
\beq
\frac{\dpar C_n(t,s)}{\dpar t}
=-C_n(t,s)+\frac{\g}{2}(C_{n-1}(t,s)+C_{n+1}(t,s))
\label{cpdot}
\eeq
for $t>s$, with the initial value
\beq
C_n(s,s)=C_n(s),
\label{cnts}
\eeq
at time $t=s$, i.e., $\tau=0$,
where the right-hand side is the equal-time correlation function,
given in Fourier-Laplace space by eqs.~(\ref{cfl}) and~(\ref{cnl}).

The second argument $s$ plays the role of a parameter in eq.~(\ref{cpdot}),
hence this equation is formally identical to eq.~(\ref{mdot}),
with the initial condition~(\ref{cnts}) playing the role of $M_n(t=0)$.
The solution of eq.~(\ref{cpdot}), seen as an evolution equation
in the $\tau$ variable, is therefore formally identical to that of
eq.~(\ref{mdot}), and reads
\beq
C_n(s+\tau,s)=C_n(s)*G_n(\tau)=\sum_m C_m(s)G_{n-m}(\tau).
\label{conv}
\eeq
In order to write down this solution explicitly,
it is convenient to introduce the double
Laplace transform of the function $C_n(s+\tau,s)$,
where $\ps$ is conjugate to $s$, the transform being denoted by $\Ls$,
and $p$ is conjugate to $\tau$, the transform being denoted by $\L$:
\beq
C_n^{\L\Ls}(p,\ps)
=\int_0^\infty\int_0^\infty C_n(s+\tau,s)\,\e^{-(\ps s+p\tau)}\,\d s\,\d\tau.
\label{defll}
\eeq
With this definition, the solution of eq.~(\ref{cpdot}) reads
\beq
C^{\F\L\Ls}(q,p,\ps)=\frac{C^{\F\L}(q,\ps)}{p+1-\g\cos q},
\label{cflla}
\eeq
i.e., using eq.~(\ref{cfl}),
\beq
C^{\F\L\Ls}(q,p,\ps)
=\frac{\sqrt{(\ps+2)^2-4\g^2}}{\ps(\ps+2-2\g\cos q)(p+1-\g\cos q)}.
\label{cfll}
\eeq

In the following, we shall mostly be interested in
the diagonal component of the correlation function,
or autocorrelation function, $C_0(t,s)$.
Eq.~(\ref{cfll}) yields, upon integration over $q$,
\beq
C_0^{\L\Ls}(p,\ps)=\frac{1}{\ps(\ps-2p)}
\left(\sqrt\frac{(\ps+2)^2-4\g^2}{(p+1)^2-\g^2}-2\right).
\label{cll}
\eeq

The discussion given at the end of section~3
on the behavior of $C_n(t)$ in various regimes
is readily extended to the two-time correlation
function $C_n(t,s)$ considered in the present section,
using its integral-transform expressions~(\ref{cfll}), (\ref{cll}).

\un{Equilibrium}

The equilibrium correlation function is obtained by letting $s\to\infty$
(in practice, $s\gg\tau_\eq)$, while keeping $\tau=t-s$ fixed.
Eq.~(\ref{cfll}) simplifies to
\beq
C_\eq^{\F\L}(q,p)=\frac{\sqrt{1-\g^2}}{(1-\g\cos q)(p+1-\g\cos q)},
\label{ceqfl}
\eeq
which yields, performing the inverse Laplace transform first,
\beq
C_{n,\eq}(\tau)=\e^{-\mu\abs{n}}-\sqrt{1-\g^2}\int_0^\tau G_n(u)\,\d u
=\sqrt{1-\g^2}\int_\tau^\infty G_n(u)\,\d u.
\label{cneqt}
\eeq
In particular the equilibrium autocorrelation function reads
\beq
C_{0,\eq}(\tau)
=\sqrt{1-\g^2}\int_\tau^\infty G_0(u)\,\d u.
\eeq
In the scaling regime,
where $\tau$ and $\tau_\eq$ are large and comparable,
using eq.~(\ref{gnscal}), the above formula simplifies to
\beq
C_{0,\eq}(\tau)
\approx\,\erfc\!\left(\sqrt\frac{\tau}{\tau_\eq}\right).
\label{ceqfull}
\eeq

\un{Non-equilibrium}

The generalization of the non-equilibrium scaling regime~(\ref{scafull})
of long times, large distances, and low temperatures,
to the present case of two temporal variables consists in taking
$s$, $\tau$, and $\tau_\eq$ simultaneously large but comparable,
with arbitrary ratios between them, and the same for $n$ and $\xi_\eq$,
i.e., using again eq.~(\ref{tauxi}),
\beq
s\sim\tau\sim\tau_\eq\sim n^2\sim\xi_\eq^2\gg1.
\label{scafull2}
\eeq
As a consequence, $p_s\sim p\sim q^2\sim\mu^2$ are simultaneously small.
In this regime, eq.~(\ref{cll}) reads
\beq
C_0^{\L\Ls}(p,\ps)\approx
\frac{2}{\ps\sqrt{2p+\mu^2}\left(\sqrt{\ps+\mu^2}+\sqrt{2p+\mu^2}\right)}.
\label{clla}
\eeq
Performing first the inverse Laplace transform of this expression
with respect to~$p$, using eq.~(A.4), we get
\beq
C_0^{\Ls}(\tau,\ps)\approx
\frac{\e^{\ps\tau/2}}{\ps}\,\erfc\left(\sqrt{(\ps+\mu^2)\tau/2}\right).
\label{c0ls}
\eeq
Performing then the inverse Laplace transform with respect to~$\ps$,
using eq.~(A.5), we obtain
\beq
C_0(s+\tau,s)\approx\frac{\sqrt{2\tau}}{\pi}\,\e^{-\mu^2\tau/2}
\int_0^s\frac{\e^{-\mu^2u}}{(2u+\tau)\sqrt{u}}\,\d u.
\eeq
The change of variable $u=(\tau/2)\tan^2\theta$ yields
\beq
C_0(s+\tau,s)\approx\frac{2}{\pi}\int_0^\Theta\,\e^{-\mu^2\tau/(2\cos^2\theta)}
\,\d\theta,
\label{ccsca}
\eeq
with
\beq
\Theta=\arctan\sqrt\frac{2s}{\tau}.
\eeq
Eq.~(\ref{ccsca}) gives the general scaling form
of the two-time autocorrelation function.
It can be alternatively expressed as a double Gaussian integral
\beq
C_0(s+\tau,s)\approx\frac{4}{\pi}\int_{\mu\sqrt{\tau/2}}^\infty\,\d\xi
\int_0^{\xi\sqrt{2s/\tau}}\e^{-(\xi^2+\eta^2)}\,\d\eta.
\eeq
Only in the symmetric situation $\Theta=\pi/4$, i.e., $\tau=2s$, or $t=3s$,
this integral simplifies to
\beq
C_0(3s,s)\approx
\frac{1}{2}\left(1-\erf\null^{\!2}\left(\mu\sqrt{s}\right)\right).
\label{dual}
\eeq
Yet another reformulation of the result~(\ref{ccsca}) reads
\beq
C_0(s+\tau,s)\approx\frac{2}{\pi}\,\sqrt\frac{2s}{\tau}\,\e^{-\mu^2\tau/2}
\,\Phi_1\!\left(\frac{1}{2},1,\frac{3}{2};-\frac{2s}{\tau},-\mu^2s\right),
\eeq
where $\Phi_1$ is the confluent hypergeometric series
in two variables~\cite{htf}, namely
the scaling variables $2s/\tau$ and $\mu^2s\approx2s/\tau_\eq$.

The \fd ratio~(\ref{dx}) involves the derivative of $C_0(t,s)$
with respect to $s$ at fixed observation time $t$, for which
eq.~(\ref{ccsca}) yields the simpler expression
\beq
\frac{\dpar C_0(t=s+\tau,s)}{\dpar s}\approx
\left(\frac{2(s+\tau)}{2s+\tau}\,\frac{\e^{-\mu^2s}}{\sqrt{\pi s}}
+\mu\,\erf(\mu\sqrt{s})\right)\frac{\e^{-\mu^2\tau/2}}{\sqrt{2\pi\tau}}.
\label{dcfull}
\eeq
Let us discuss some limiting cases of eqs.~(\ref{ccsca}) and~(\ref{dcfull}).

\noindent(i)
At zero temperature, and more generally in the domain-growth regime~(\ref{l2}),
the autocorrelation function exhibits aging,
i.e., it only depends on the ratio $t/s$.
Eq.~(\ref{ccsca}) indeed simplifies
for $\mu=0$ to $C_0(t,s)\approx2\Theta/\pi$, i.e.,
\beq
C_0(s+\tau,s)\approx\frac{2}{\pi}\arctan\sqrt\frac{2s}{\tau}.
\label{csca}
\eeq
This expression~\cite{bray2,amar,bray3,PBS} assumes the scaling form
\beq
C_0(s+\tau,s)\approx F(x),
\label{cscaf}
\eeq
with
\beq
x=\frac{t}{s}=1+\frac{\tau}{s}\ge1,
\label{defx}
\eeq
and
\beq
F(x)=\frac{2}{\pi}\arctan\sqrt\frac{2}{x-1}.
\label{fx}
\eeq
We have $F(3)=1/2$, in agreement with eq.~(\ref{dual}).
Similarly, eq.~(\ref{dcfull}) becomes
\beq
\frac{\dpar C_0(t,s)}{\dpar s}\approx\frac{F_1(x)}{s},
\label{dcsca}
\eeq
with $F_1(x)=-x\,\d F(x)/\d x$, i.e.,
\beq
F_1(x)=\frac{x}{\pi(x+1)}\sqrt\frac{2}{x-1}.
\label{f1x}
\eeq

\noindent(ii)
For $s\to\infty$ with $\tau$ fixed, i.e., $p_s\to0$, eq.~(\ref{ccsca})
converges to the equilibrium form~(\ref{ceqfull});
this is obvious using eq.~(\ref{c0ls}).

\noindent(iii)
The early-time regime, or temporal Porod regime~(\ref{l3}),
is the counterpart of the spatial Porod regime~(\ref{porods}).
From eq.~(\ref{dcfull}), we find
\beq
\frac{\dpar C_0(s+\tau,s)}{\dpar s}\approx\frac{A(s)}{\sqrt{2\pi\tau}},
\label{dcas}
\eeq
where $A(s)$ has been defined in eq.~(\ref{as}).
We thus obtain, to leading order,
\beq
C_0(s+\tau,s)\approx 1-A(s)\sqrt\frac{2\tau}{\pi}.
\label{cporodt}
\eeq
The replacement of $\abs{n}$ in eq.~(\ref{cporods})
by (a constant times) $\sqrt\tau$ in eq.~(\ref{cporodt})
reflects once more the underlying diffusive mechanism.
Finally, the behavior~(\ref{adom}) of $A(s)$ in the domain-growth regime
is in agreement with eq.~(\ref{csca}),
while eq.~(\ref{aeq}) matches eq.~(\ref{ceqfull}).
The generalization to the spatio-temporal Porod regime
\beq
1\ll n^2\sim\tau\ll s\sim\tau_\eq
\eeq
is straightforward.
One has, using eqs.~(\ref{conv}), (\ref{cporods}), (\ref{gnscal}),
\beq
C_n(s+\tau,s)\approx(1-A(s)|n|)
*\frac{\e^{-n^2/(2\tau)}}{\sqrt{2\pi\tau}},
\eeq
where the convolution involves the spatial coordinate $n$.
Replacing the discrete sum by an integral, we obtain
\beq
C_n(s+\tau,s)\approx 1-A(s)\left\{\sqrt\frac{2\tau}{\pi}\,\e^{-n^2/(2\tau)}
+n\,\erf\left(\frac{n}{\sqrt{2\tau}}\right)\right\}.
\label{cporodst}
\eeq
This result interpolates between the spatial behavior~(\ref{cporods})
of the equal-time correlation function, for $\tau=0$,
and the temporal behavior~(\ref{cporodt}) of the autocorrelation
function, for $n=0$.

\section{Two-time response function}

Suppose now that the system is subjected to a small magnetic field $H_n(t)$,
depending on the site label $n$ and on time $t>0$ in an arbitrary fashion.
This amounts to adding to the ferromagnetic Hamiltonian~(\ref{ham})
a time-dependent perturbation of the form
\beq
\delta\H(t)=-\sum_n H_n(t)\s_n(t).
\eeq
The dynamics of the model is still given by the stochastic rule~(\ref{update}),
where the local field $h_n(t)$ acting on the spin $\s_n(t)$ now reads
\beq
h_n(t)=J(\s_{n-1}(t)+\s_{n+1}(t))+H_n(t).
\eeq

We again consider a random initial condition.
Causality and invariance under spatial translations imply that
the magnetization $M_n(t)$ at time $t$ reads
\beq
M_n(t)=\mean{\s_n(t)}=\beta\int_0^t\d u\sum_m R_{n-m}(t,u)H_m(u)+\cdots,
\eeq
to first order in the magnetic fields $H_n(t)$.
This formula defines the
two-time dimensionless response function $R_{n-m}(t,s)$ of the model.
A more formal definition reads
\beq
R_{n-m}(t,s)=T\left.\frac{\delta M_n(t)}{\delta H_m(s)}\right\vert_{\{H=0\}}.
\eeq

The evolution equation~(\ref{mvt}) still holds
in the presence of arbitrary external magnetic fields.
Furthermore we have, to first order in the magnetic field $H_n$,
\beq
\tanh(\beta h_n)=\tanh(\beta J(\s_{n-1}+\s_{n+1}))
+\beta H_n\left(1-\tanh^2(\beta J(\s_{n-1}+\s_{n+1}))\right)+\cdots,
\eeq
together with the identities
\beqa
\tanh(\beta J(\s_{n-1}+\s_{n+1}))
&=&\frac{\g}{2}(\s_{n-1}+\s_{n+1}),\nonumber\\
\tanh^2(\beta J(\s_{n-1}+\s_{n+1}))
&=&\frac{\g^2}{2}(1+\s_{n-1}\s_{n+1}).
\eeqa
Inserting these expressions into eq.~(\ref{mvt}), we readily obtain that,
again to first order in the magnetic field $H_n(t)$,
the magnetizations $M_n(t)$ obey inhomogeneous
differential equations of the form
\beq
\frac{\d M_n(t)}{\d t}=-M_n(t)+\frac{\g}{2}(M_{n-1}(t)+M_{n+1}(t))
+\beta H_n(t)\left(1-\frac{\g^2}{2}(1+C_{n-1,n+1}(t))\right)+\cdots
\label{mdoth}
\eeq

As a consequence, the two-time response function $R_n(t,s)$ itself
obeys coupled linear differential equations, of the form
\beq
\frac{\dpar R_n(t,s)}{\dpar t}
=-R_n(t,s)+\frac{\g}{2}(R_{n-1}(t,s)+R_{n+1}(t,s)),
\label{rdot}
\eeq
for $t>s$, with the initial value
\beq
R_n(s,s)=w(s)\delta_{n,0},
\label{rnss}
\eeq
and with
\beq
w(s)=1-\frac{\g^2}{2}(1+C_2(s)),
\label{ws}
\eeq
where $C_2(s)$ is the equal-time correlation two sites apart,
which is given in Laplace space by eq.~(\ref{cnl}).
Eq.~(\ref{rdot}), with its initial condition~(\ref{rnss}),
is formally identical to eq.~(\ref{cpdot}) with initial condition~(\ref{cnts}).
Hence its solution reads
\beq
R_n(s+\tau,s)=w(s) G_n(\tau),
\label{rnts}
\eeq
or, in Laplace space with respect to $s$,
\beq
R_n^{\Ls}(\tau,\ps)=w^{\Ls}(\ps)G_n(\tau).
\label{rfll}
\eeq
Using eq.~(\ref{cnl}), we obtain
\beq
w^{\Ls}(\ps)=\frac{\ps+2}{4\ps}\sqrt{(\ps+2)^2-4\g^2}-1-\frac{\ps}{4}.
\label{wl}
\eeq

Let us now discuss the general expression of the response function,
given in Laplace space by eqs.~(\ref{rfll}) and~(\ref{wl}),
along the same lines as for the two-time correlation function.

\un{Equilibrium}

At equilibrium, i.e., for $s\to\infty$ with $\tau$ fixed,
one gets, either using eq.~(\ref{cneq})
or taking the $p_s\to0$ limit of eq.~(\ref{wl}),
\beq
w_\eq=\sqrt{1-\g^2},
\eeq
hence
\beq
R_{n,\eq}(\tau)=\sqrt{1-\g^2}\, G_n(\tau).
\label{reqfl}
\eeq
Comparing this equation to eq.~(\ref{cneqt}) leads to the identity
\beq
R_{n,\eq}(\tau)=-\frac{\d C_{n,\eq}(\tau)}{\d\tau},
\eeq
which is the \fd theorem~(\ref{fdt}), in dimensionless form.

\un{Non-equilibrium}

For simplicity we again focus our attention on the diagonal component
$R_0(t,s)$.
In the scaling regime~(\ref{scafull2}), we have
\beq
w^{\Ls}(\ps)=\frac{\sqrt{p_s+\mu^2}}{p_s}.
\label{wl2}
\eeq
Eq.~(A.2) implies that the function $w(s)$ coincides
in this regime with the amplitude $A(s)$, defined in eq.~(\ref{as}).
Using eq.~(\ref{rfll}), and the scaling form~(\ref{gnscal}) of $G_0(\tau)$,
we obtain the remarkably simple result
\beq
R_0(s+\tau,s)\approx\frac{A(s)\,\e^{-\mu^2\tau/2}}{\sqrt{2\pi\tau}}.
\label{rfull}
\eeq
This general scaling form of the two-time response function
further simplifies in the following limiting cases.

\noindent(i)
At zero temperature, and more generally in the aging regime~(\ref{l2}),
the result~(\ref{rfull}) becomes
\beq
R_0(s+\tau,s)\approx\frac{1}{\pi\sqrt{2s\tau}}.
\label{rpro}
\eeq
We thus obtain a scaling form similar to eq.~(\ref{dcsca}), namely
\beq
R_0(s+\tau,s)\approx\frac{F_2(x)}{s},
\label{rsca}
\eeq
with
\beq
F_2(x)=\frac{1}{\pi\sqrt{2(x-1)}}.
\label{f2x}
\eeq

\noindent(ii)
In the early-time regime~(\ref{l3}), we obtain
\beq
R_0(s+\tau,s)\approx\frac{A(s)}{\sqrt{2\pi\tau}},
\label{ras}
\eeq
where the right-hand side is identical to eq.~(\ref{dcas}).

\section{Fluctuation-dissipation ratio}

As mentioned in the introduction, a way of characterizing the violation of
the \fd theorem~(\ref{fdt}) in a non-equilibrium situation
consists in introducing the \fd ratio $X(t,s)$ of eq.~(\ref{dx}).
In the present situation, we set
\beq
X(t,s)=\frac{R_0(t,s)}{\,\frad{\dpar C_0(t,s)\haut}{\dpar s}\,},
\label{dx2}
\eeq
because the response function $R_0(t,s)$ is dimensionless.

In the non-equilibrium scaling regime~(\ref{scafull2}),
eqs.~(\ref{dcfull}) and~(\ref{rfull}) yield
\beq
X(s+\tau,s)\approx
\frac{\e^{-\mu^2s}+\mu\sqrt{\pi s}\,\erf(\mu\sqrt{s})}
{\frad{2(s+\tau)\haut}{2s+\tau}\,\e^{-\mu^2s}+\mu\sqrt{\pi
s}\,\erf(\mu\sqrt{s})},
\label{xfull}
\eeq
which is the scaling form of the \fd ratio in the variables
$s/\tau_\eq$ and $\tau/\tau_\eq$.
We again discuss some limiting cases.

\noindent(i)
For $s\gg\tau_\eq$, with $\tau$ and $\tau_\eq$ fixed,
the \fd ratio converges exponentially fast to its equilibrium value $X_\eq=1$,
according to
\beq
X(s+\tau,s)\approx
1-\frac{\tau}{2s}\sqrt{\frac{\tau_\eq}{2\pi s}}\,\e^{-2s/\tau_\eq}.
\label{cvx}
\eeq

\noindent(ii)
The result~(\ref{xfull}) also shows that we have $X(t,s)\approx1$
in the early-time regime~(\ref{l3}), consistently with the identity
between expressions~(\ref{dcas}) and~(\ref{ras}).
Notice, however, that the initial value $X(s,s)$ of the \fd ratio
is not identically equal to~1.
For instance, at zero temperature, eqs.~(\ref{cll}) and~(\ref{wl})
yield after some algebra
\beq
X(s,s)=1-\frac{I_1(2s)}{2s(I_0(2s)+I_1(2s)\haut)}.
\eeq
This expression behaves as $X(s,s)\approx1/2$
in the regime $s\ll1$ of no physical interest,
while it converges to unity after a microscopic transient regime,
as $X(s,s)\approx1-1/(4s)$ for $s\gg1$.

\noindent(iii)
At zero temperature, and more generally in the aging regime~(\ref{l2}),
eq.~(\ref{xfull}) simplifies to
\beq
X(s+\tau,s)\approx\X(x),
\label{xsca}
\eeq
where the scaling function $\X(x)$ reads
\beq
\X(x)=\frac{x+1}{2x}.
\label{phix}
\eeq
This result is consistent with the
scaling laws~(\ref{dcsca}) and~(\ref{rsca}), as we have
\beq
\X(x)=\frac{F_2(x)}{F_1(x)}.
\eeq
The \fd ratio decreases from the initial value $\X(0)=1$,
corresponding to equilibrium, to the non-trivial asymptotic value
\beq
X_\infty=\X(\infty)=1/2.
\label{xuni}
\eeq

Recent developments on aging systems~\cite{revue} suggest the following
alternative presentations of the above results concerning the aging regime.

Firstly, there is a functional relationship $X(C)$
between the \fd ratio $X\equiv X(t,s)$ and
the two-time correlation function $C\equiv C_0(t,s)$ in the aging regime,
as a consequence of eqs.~(\ref{cscaf}) and~(\ref{xsca}).
Indeed, eliminating the time ratio $x$ between eqs.~(\ref{fx})
and~(\ref{phix}), we obtain
\beq
X(C)=\frac{1}{1+\cos^2\!\left(\frad{\pi C\haut}{2}\right)}.
\label{xc}
\eeq

\vskip 8.5cm{\hskip 0.8cm}
\includegraphics{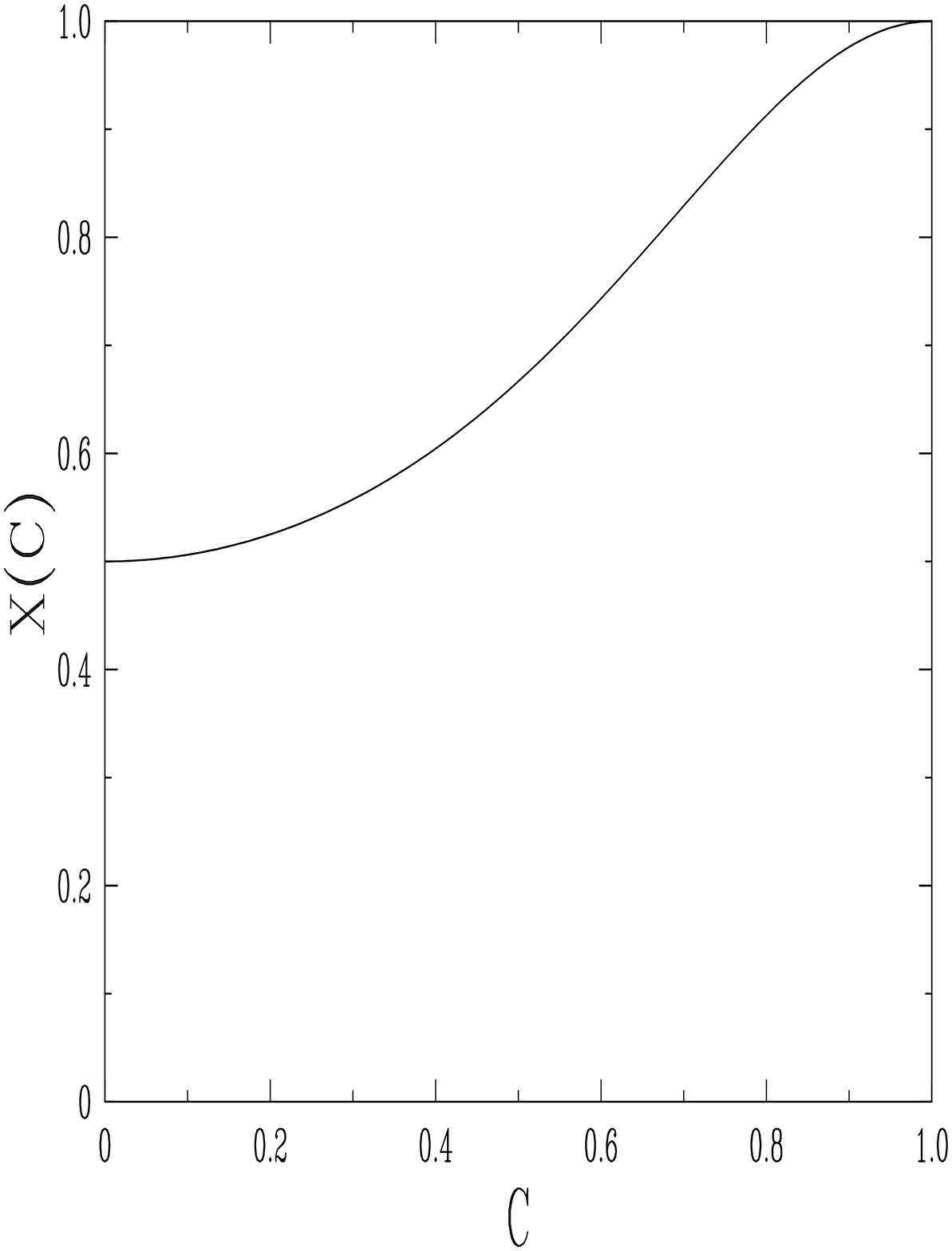}

\noindent{\bf Figure~1:} Plot of the relationship $X(C)$ of eq.~(\ref{xc})
between \fd ratio $X(t,s)$ and correlation function $C_0(t,s)$
in the aging regime.

\vskip .3cm

Secondly, the dimensionless integrated response function
\beq
\R_0(t,s)=\int_0^s R_0(t,u)\,\d u,
\eeq
proportional to the thermoremanent magnetization~\cite{revue}, reads
\beq
\R_0(t,s)=\int_{C_0(t,0)}^{C_0(t,s)}X(t,u)\,\d C_0(t,u),
\eeq
using the definition~(\ref{dx2}) of $X(t,s)$.
In the aging regime, the existence of the functional
relationship~(\ref{xc}) implies
\beq
\R_0(t,s)\approx\R(C_0(t,s)),
\eeq
with
\beq
\R(C)=\int_0^C X(C')\,\d C',
\eeq
i.e., explicitly,
\beq
\R(C)=\frac{\sqrt2}{\pi}
\arctan\left(\frac{1}{\sqrt2}\,\tan\left(\frad{\pi C}{2}\right)\right).
\label{rc}
\eeq

\vskip 8.5cm{\hskip 0.8cm}
\includegraphics{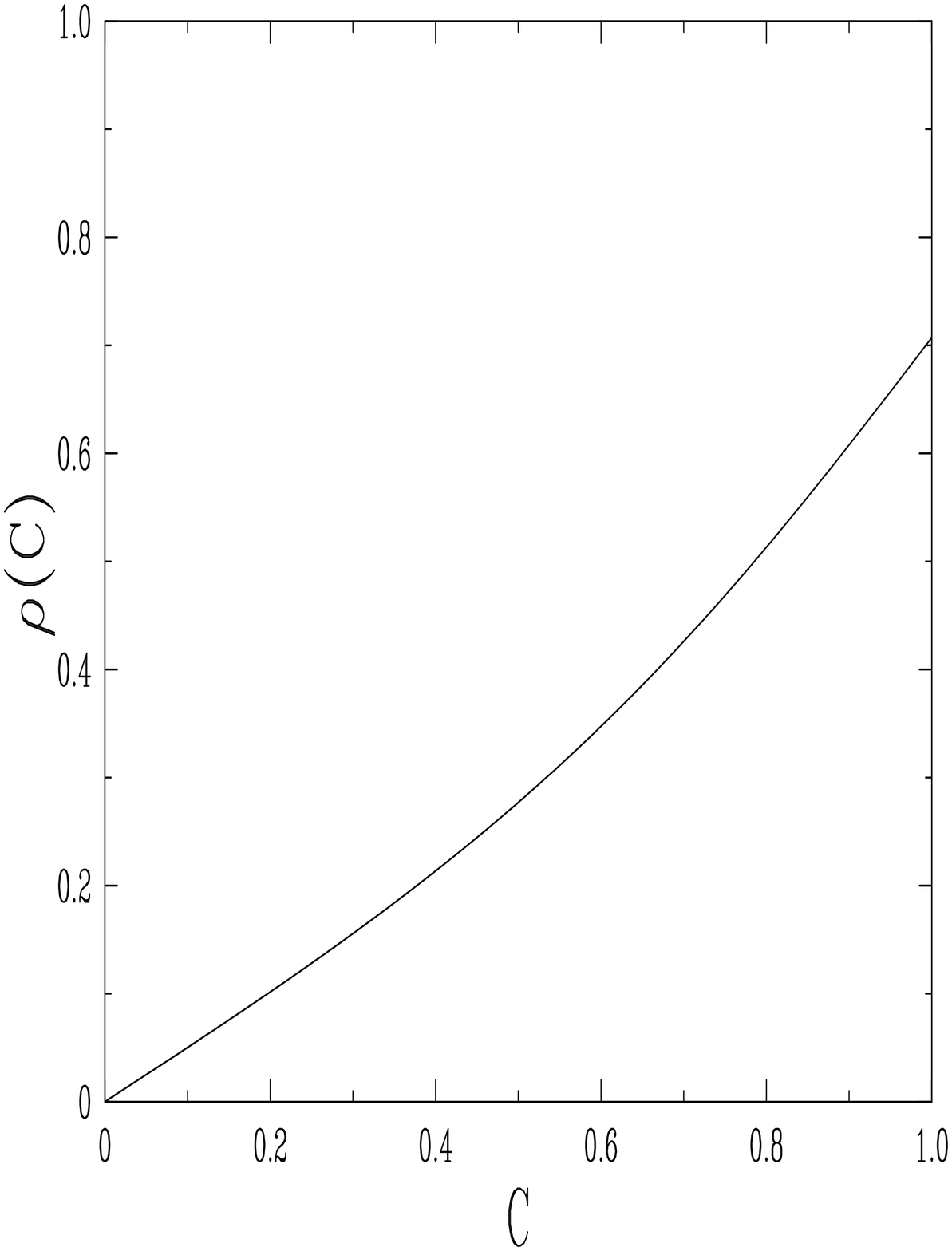}

\noindent{\bf Figure~2:} Plot of the relationship $\R(C)$ of eq.~(\ref{rc})
between integrated response $\R_0(t,s)$ and correlation function $C_0(t,s)$
in the aging regime.

\vskip .3cm

The functions $X(C)$ and $\R(C)$ are respectively shown in Figures~1 and~2.
For $C\to1$, i.e., $x=t/s\to1$, we have
\beq
X(C)=1-\frac{\pi^2}{4}(1-C)^2+\cdots,\qquad
\R(C)=\frac{1}{\sqrt2}-(1-C)+\frac{\pi^2}{12}(1-C)^3+\cdots,
\eeq
while for $C\to0$, i.e., $x\to\infty$, we have
\beq
X(C)=\frac{1}{2}+\frac{\pi^2C^2}{16}+\cdots,\qquad
\R(C)=\frac{C}{2}+\frac{\pi^2C^3}{48}+\cdots
\eeq
Let us point out that the slope of the $\R(C)$ curve near the origin
is given by the asymptotic value $X(C=0)=X_\infty=1/2$.

It is clear from eq.~(\ref{xfull}) that the \fd theorem is maximally violated
for $\tau\gg s$, at least in the scaling regime~(\ref{scafull}).
In this situation, eq.~(\ref{xfull}) yields the prediction
\beq
X_\as(s)\approx\frac{\e^{-\mu^2s}+\mu\sqrt{\pi s}\,\erf(\mu\sqrt{s})}
{2\e^{-\mu^2s}+\mu\sqrt{\pi s}\,\erf(\mu\sqrt{s})}
\label{xas}
\eeq
for the asymptotic \fd ratio
\beq
X_\as(s)=\lim_{\tau\to\infty}X(s+\tau,s).
\eeq

\vskip 8.5cm{\hskip 0.8cm}
\includegraphics{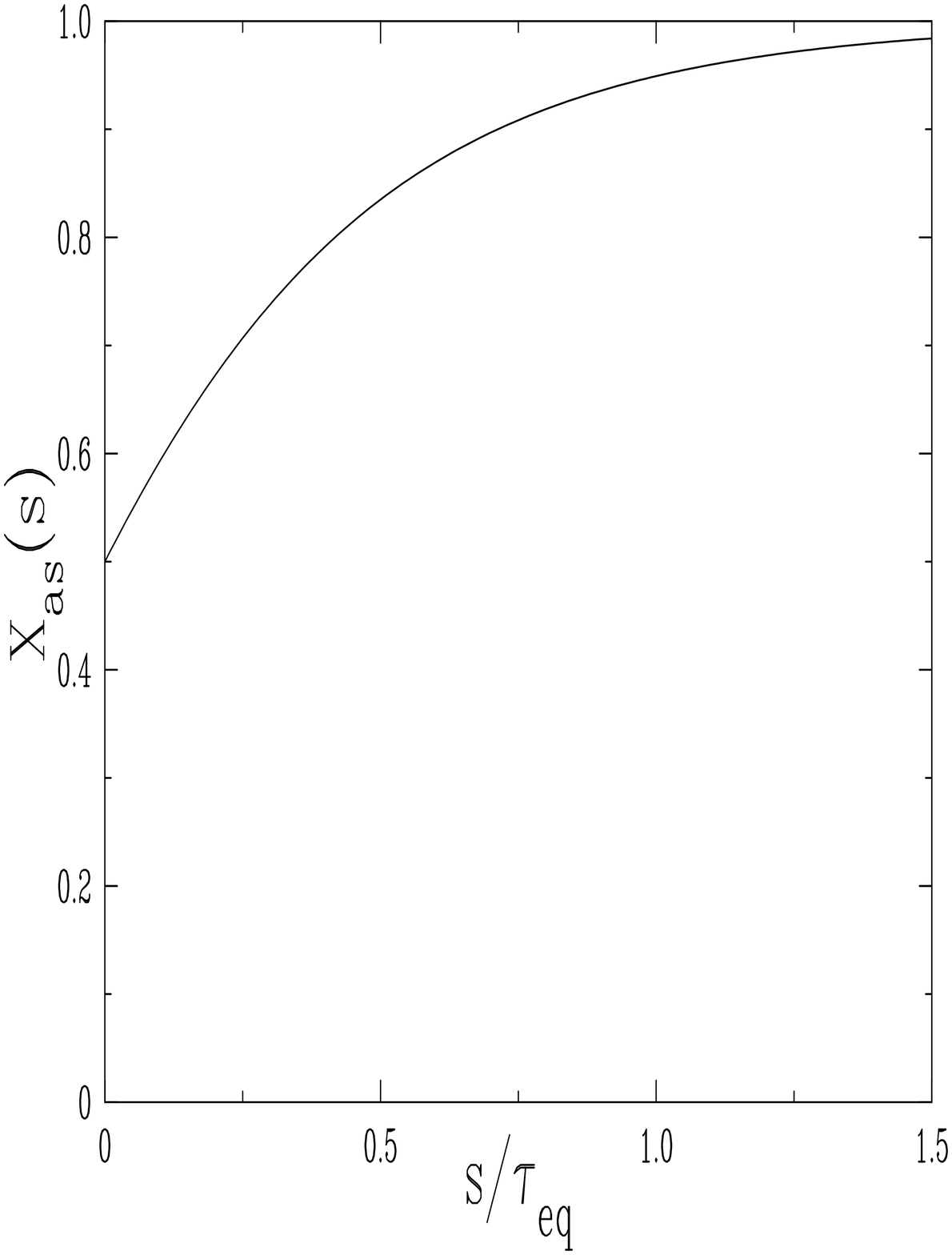}

\noindent{\bf Figure~3:} Plot of the asymptotic \fd ratio $X_\as(s)$
in the non-equi\-li\-bri\-um scaling regime,
as given by eq.~(\ref{xas}), against $s/\tau_\eq$.

\vskip .3cm

Figure~3 shows a plot of the prediction~(\ref{xas}) for $X_\as(s)$,
against the ratio $s/\tau_\eq$.
In the aging regime $(1\ll s\ll\tau_\eq)$, the asymptotic \fd ratio
smoothly departs from its limit value $X_\infty=1/2$, as
\beq
X_\as(s)\approx\frac{1}{2}+\frac{s}{\tau_\eq}+\cdots,
\eeq
while it converges exponentially fast to its equilibrium value $X_\eq=1$
for $s\gg\tau_\eq$:
\beq
X_\as(s)\approx1-\sqrt{\frac{\tau_\eq}{2\pi s}}\,\e^{-2s/\tau_\eq}.
\label{cvxas}
\eeq

\section{Discussion}

This work is devoted to the non-equilibrium dynamics
of the ferromagnetic Ising chain,
quenched from infinite temperature to finite temperature,
and evolving under Glauber dynamics.
We have exploited the solvability of this model
in order to derive exact expressions for the spin autocorrelation function
$C_0(t,s)=\mean{\s_0(t)\s_0(s)}$,
and the associated response function $R_0(t,s)$ and \fd ratio $X(t,s)$,
with $t$ (observation time) $>s$ (preparation time) and $\tau=t-s$.
While our study of correlations complements a well-investigated field,
the results concerning the response and the \fd ratio are entirely novel.

We have given a detailed analysis of the scaling regime of low temperatures,
large distances (proportional to the equilibrium correlation length $\xi_\eq$)
and large times (proportional to the relaxation time $\tau_\eq\sim\xi_\eq^2$).
This scaling regime encompasses several limiting situations of interest:

\un{
Self-similar domain-growth or aging regime $(1\ll s\sim\tau\ll\tau_\eq)$.}
At zero temperature, because of the self-similarity
of the coarsening phenomenon,
the various two-time observables obey simple scaling laws
involving only the time ratio $x=t/s$.
This situation is typical of an aging system~\cite{revue}.
The corresponding scaling functions for the autocorrelation function
$C_0(t,s)$, its derivative $\dpar C_0(t,s)/\dpar s$,
the response function $R_0(t,s)$, and the associated
\fd ratio $X(t,s)$ are given explicitly in eqs.~(\ref{fx}),
(\ref{f1x}), (\ref{f2x}), and (\ref{phix}).
These functions are universal,
implying in particular that they are unchanged if the initial condition
contains short-range correlations.
These results imply the existence of a non-trivial relationship
$X(C)$ [see eq.~(\ref{xc})] throughout the aging regime.
At any low but finite temperature,
the self-similar domain growth and the associated aging phenomena
are interrupted for times of order $\tau_\eq$.

\un{
Spatial $(1\ll\abs{n}\ll\xi_\eq\sim\sqrt{s})$
and temporal $(\tau\ll s\sim\tau_\eq)$ Porod regimes.}
In these regimes, the autocorrelation function departs
from its value of unity in a singular fashion,
involving either an $\abs{n}$~\cite{bray1} or a $\sqrt{\tau}$
dependence~[see eqs.~(\ref{cporods}), (\ref{cporodt})].
We have derived a prediction~(\ref{as})
for the common prefactor $A(t)$ of both these laws,
as well as the interpolation formula~(\ref{cporodst})
in the spatio-temporal regime.

\un{
Equilibrium regime $(\tau\sim\tau_\eq\ll s)$.}
At low but finite temperature, the system converges
toward equilibrium for times larger than $\tau_\eq$.
The exponential law of convergence has also been studied
in detail~[see eqs.~(\ref{cvcn}), (\ref{cvx}), (\ref{cvxas})].

As already underlined in the introduction,
one of the most salient outcomes of this study
is the non-trivial limit value $X_\infty=1/2$ of the \fd ratio
for $1\ll s\ll t$ in the aging regime.
The value of $X_\infty$ for a system quenched to its critical point
could be a priori any number between
$X_\infty=1$ ($T>T_c$: equilibrium) and $X_\infty=0$ ($T<T_c$: domain growth).
It turns out that the answer
is exactly half-way between these bounds for the Glauber-Ising chain.
A forthcoming work~\cite{ustocome}
is devoted to the limit \fd ratio in systems having a finite $T_c$,
and exhibiting domain growth in a whole low-temperature phase.
The spherical model in any dimension $d>2$
and the two-dimensional Ising model will serve as benchmarks
and lead us to claim that $X_\infty$ is a novel universal characteristic
of critical dynamics, intrinsically related to non-equilibrium phenomena,
which can take any value, at least in the range $0\le X_\infty\le1/2$.

Let us anticipate that the limit \fd ratio $X_\infty$ appears as
an amplitude ratio~\cite{ustocome},
in the sense used in critical phenomena.
For a critical quench in such generic models, we have indeed
\beqa
C(t,s)&\approx&s^{-(d-2+\eta)/z}\,F(x),\cr
\frac{\dpar C\haut(t,s)}{\dpar\bas s}&\approx&s^{-1-(d-2+\eta)/z}\,F_1(x),\cr
T_c\,R(t,s)&\approx&s^{-1-(d-2+\eta)/z}\,F_2(x).
\label{scagen}
\eeqa
These scaling laws generalize eqs.~(\ref{cscaf}), (\ref{dcsca}), (\ref{rsca}).
The dynamical exponent $z$ is
accessible from the study of dynamical critical phenomena at equilibrium,
while $x=t/s$ is again the time ratio.
The \fd ratio therefore scales as
\beq
X(t,s)\approx\X(x)=\frac{F_2(x)}{F_1(x)},
\eeq
where the scaling function $\X(x)$ is again universal.
For large values of this ratio $(x\gg1$, i.e., $1\ll s\ll t)$,
the scaling functions entering eqs.~(\ref{scagen}) fall off as
\beqa
F(x)&\approx&B\,x^{-\lambda_c/z},\cr
F_1(x)&\approx&B_1\,x^{-\lambda_c/z},\cr
F_2(x)&\approx&B_2\,x^{-\lambda_c/z},
\label{f12}
\eeqa
where $\lambda_c$, related to the critical initial-slip exponent
$\theta_c$~\cite{janssen} by $\lambda_c=d-z\theta_c$,
is a dynamical critical exponent
which only manifests itself in non-equilibrium phenomena~\cite{huse}.
The exact results derived in this paper for the Ising chain
agree with the above scaling laws
for $\eta=1$, $z=2$, $\lambda_c=1$, and $\theta_c=0$.
The limit \fd ratio
\beq
X_\infty=\X(\infty)=\frac{B_2}{B_1}
\eeq
thus appears as a dimensionless amplitude ratio
associated with the behavior~(\ref{f12}).
It is therefore a novel universal quantity
in non-equilibrium critical dynamics.

\newpage
\appendix
\section{Some useful inverse Laplace transforms}

In this Appendix we list a few inverse Laplace transforms,
which can be found in ref.~\cite{tit},
and have been used in the body of this article.
Notations are as in eq.~(\ref{lapdef}).
The symbols $a$, $b$ denote complex parameters with a positive real part.

\[
\begin{array}{llr}
&&\\
f^\L(p)&f(t)&\\
&&\\
\frad{2\,\e^{-b\sqrt{p+a^2}}}{p}&
\e^{ab}\,\erfc\!\left(\frad{b}{2\sqrt{t}}+a\sqrt{t}\right)
+\e^{-ab}\,\erfc\!\left(\frad{b}{2\sqrt{t}}-a\sqrt{t}\right)
{\hskip .4cm}&(\A.1)\\
&&\\
\frad{\sqrt{p+a^2}}{p}
&\frad{\e^{-a^2t}}{\sqrt{\pi t}}+a\,\erf\left(a\sqrt{t}\right)&(\A.2)\\
&&\\
\frad{a}{p\sqrt{p+a^2}}&\erf\left(a\sqrt{t}\right)&(\A.3)\\
&&\\
\frad{1}{\sqrt{p+a^2}\left(b+\sqrt{p+a^2}\right)}{\hskip .66cm}
&\e^{(b^2-a^2)t}\,\erfc\left(b\sqrt{t}\right)&(\A.4)\\
&&\\
\e^{ap}\,\erfc\left(\sqrt{ap}\right)
&\frad{1}{\pi(a+t)}\sqrt\frad{a}{t}&(\A.5)\\
\end{array}
\nonumber
\]


\newpage
\begin{thebibliography}{99}

\bibitem{glauber} R.J. Glauber, J. Math. Phys. {\bf 4}, 294 (1963).

\bibitem{langer} J.S. Langer, in {\it Solids far from Equilibrium}, edited by
C. Godr\`eche (Cambridge University Press, 1991).

\bibitem{bray1} A.J. Bray, Adv. Phys. {\bf 43}, 357 (1994).

\bibitem{revue} For recent reviews, see: E. Vincent, J. Hammann, M. Ocio,
J.P. Bouchaud, and L.F. Cugliandolo, preprint cond-mat/9607224,
published in {\it Complex Behavior of Glassy Systems},
Springer Lecture Notes in Physics {\bf 492}, 184 (1997);
J.P. Bouchaud, L.F. Cugliandolo, J. Kurchan, and M.~M\'ezard,
preprint cond-mat/9702070, published in {\it Spin Glasses and Random Fields},
Directions in Condensed Matter Physics, vol.~{\bf 12},
edited by A.P. Young (World Scientific, Singapore, 1998).

\bibitem{ckp} L.F. Cugliandolo, J. Kurchan, and G. Parisi, J. Phys. I (France)
{\bf 4}, 1641 (1994).

\bibitem{ck} L.F. Cugliandolo and J. Kurchan, J. Phys. A {\bf 27}, 5749 (1994).

\bibitem{x1} L.F. Cugliandolo, J. Kurchan, and L. Peliti, Phys. Rev. E
{\bf 55}, 3898 (1997).

\bibitem{x2} A. Barrat, Phys. Rev. E {\bf 57}, 3629 (1998);
L. Berthier, J.L. Barrat, and J. Kurchan, preprint cond-mat/9903091.

\bibitem{autres} S. Franz, M. M\'ezard, G. Parisi, and L. Peliti, preprint
cond-mat/9903370.

\bibitem{fr97} S. Franz and F. Ritort, J. Phys. A {\bf 30}, L 359 (1997),
and references therein.

\bibitem{gl99} C. Godr\`eche and J.M. Luck, J. Phys. A {\bf 32}, 6033 (1999),
and references therein.

\bibitem{cox} J.T. Cox and D. Griffeath, Ann. Probab. {\bf 14}, 347 (1986).

\bibitem{bray2} A.J. Bray, J. Phys. A {\bf 22}, L 67 (1989).

\bibitem{amar} J.G. Amar and F. Family, Phys. Rev. A {\bf 41}, 3258 (1990).

\bibitem{bray3} A.J. Bray, in {\it Nonequilibrium Statistical Mechanics
in One Dimension}, edited by V.~Privman (Cambridge University Press, 1997).

\bibitem{PBS} A. Prados, J.J. Brey, and B. S\'anchez-Rey, Europhys. Lett.
{\bf 40}, 13 (1997).

\bibitem{ustocome} C. Godr\`eche and J.M. Luck, in preparation.

\bibitem{baxter} R.J. Baxter, {\it Exactly Solved Models in Statistical
Mechanics} (Academic Press, London, 1982).

\bibitem{htf} A. Erd\'elyi (ed.), {\it Higher Transcendental Functions}
(McGraw-Hill, New-York, 1953).

\bibitem{janssen} H.K. Janssen, B. Schaub, and B. Schmittmann, Z. Phys. B
{\bf 73}, 539 (1989).

\bibitem{huse} D.A. Huse, Phys. Rev. B {\bf 40}, 304 (1989).

\bibitem{tit} A. Erd\'elyi (ed.), {\it Tables of Integral Transforms}
(McGraw-Hill, New-York, 1954).

\end{thebibliography}
\end{document}